# Resonant interaction of the electron beam with a synchronous wave in controlled magnetrons for high-current superconducting accelerators

G. Kazakevich[*] and R. Johnson
*Muons, Inc., Batavia, Illinois 60510, USA*

V. Lebedev and V. Yakovlev
*Fermilab, Batavia, Illinois 60510, USA*

V. Pavlov
*Budker Institute of Nuclear Physics, Novosibirsk 630090, Russia*

A simplified analytical model of the resonant interaction of the beam of Larmor electrons drifting in the crossed constant fields of a magnetron with a synchronous wave providing a phase grouping of the drifting charge was developed to optimize the parameters of an rf resonant injected signal driving the magnetrons for management of phase and power of rf sources with a rate required for superconducting high-current accelerators. The model, which considers the impact of the rf resonant signal injected into the magnetron on the operation of the injection-locked tube, substantiates the recently developed method of fast power control of magnetrons in the range up to 10 dB at the highest generation efficiency, with low noise, precise stability of the carrier frequency, and the possibility of wideband phase control. Experiments with continuous wave 2.45 GHz, 1 kW microwave oven magnetrons have verified the correspondence of the behavior of these tubes to the analytical model. A proof of the principle of the novel method of power control in magnetrons, based on the developed model, was demonstrated in the experiments. The method is attractive for high-current superconducting rf accelerators. This paper also discusses vector methods of power control with the rates required for superconducting accelerators, the impact of the rf resonant signal injected into the magnetron on the rate of phase control of the injection-locked tubes, and a conceptual scheme of the magnetron transmitter with highest efficiency for high-current accelerators.

## I. INTRODUCTION

The rf sources feeding superconducting rf (SRF) cavities in modern accelerators require wideband continuous control of the phase and power to maintain instability of the accelerating field phase and amplitude about of 0.1° and 0.1%, respectively, or less. This allows one to compensate for parasitic modulations inherent in superconducting cavity operation. Depending on the SRF cavity type and mode of operation, its fundamental mode bandwidth may be of the same order as the bandwidth of mechanical oscillations of the cavity walls caused by microphonics, Lorentz force detuning (LFD), etc. [1,2]. This causes sufficiently large amplitude and phase deviations of the accelerating field, which may vary from cavity to cavity. Compensation of the parasitic amplitude deviations by an active electronic damping [3] requires an individual variation of the rf power feeding each SRF cavity that can be quite large. Traditionally, rf amplifiers such as klystrons, inductive output tubes, or solid-state amplifiers are used as high-power sources. They provide power levels up to hundreds of kW or more in continuous wave (CW) mode at a carrier frequency in the GHz range with a bandwidth in the MHz range allowing compensation of the modulations. However, the capital cost for a unit of power for the traditional rf sources is quite high, at least a few dollars per Watt [4]. Therefore, when utilizing traditional rf sources for large-scale accelerator facilities that shall deliver beam power levels of several MWs—such as next generation neutron sources or accelerator driven systems (ADS) for subcritical reactors, etc.; the capital cost of the rf system is a significant fraction of the overall accelerator project cost. In contrast, the cost of a unit of power of a commercial, L-band, CW, high-power magnetron rf source is several times less [4]. Since the magnetron efficiency is higher compared to traditional rf sources, magnetron transmitters will allow a significant reduction of both the capital and operating costs in large-scale accelerator projects.

[*]Present address: Fermi National Accelerator Laboratory, P.O. Box 500, Batavia, Illinois 60510, USA.
gkazakevitch@yahoo.com

Phase-locking, i.e., stabilization of the magnetron phase without an active control, was considered for rf accelerators in a number of works in the past with the assumption that the magnetrons were the main sources of the phase instability. The consideration was based on the study of the locking phenomena in a triode oscillator [5]. Later, magnetrons were represented as amplifiers [6,7], and the concept of injection locking was applied to them. The phase locking of magnetrons to synchronize them in normal conducting accelerators was suggested in [8]. In [9], the authors utilized the injection locking of a magnetron by the wave reflected from an accelerating cavity. A transient process in the magnetron-accelerating cavity system was considered in this article by the abridged equations. The technique provided stability of the accelerating frequency sufficient for operation of the 1–3 THz free electron laser (FEL) with a microtron-injector driven by the magnetron.

In superconducting accelerators, the above-mentioned parasitic modulations are not caused by the instability of rf sources; they are inherent in SRF cavity operation. Thus, powering SRF cavities requires locking of the phase and amplitude of the accelerating field [10] with an rf source that must be dynamically managed in phase and power with an appropriate bandwidth of the control. The phase stabilization of the accelerating field in a 2.45 GHz SRF cavity by a dynamically controlled phase-modulated resonant injected signal, providing an rms phase deviation of $\approx 1°$, was first demonstrated in Refs. [11,12].

A fast method of phase control in magnetrons is realized by a wideband phase modulation of the driving resonant (injection-locking) rf signal [13]. Presently three methods have been suggested and tested to change the power of the magnetron based transmitter in accordance with the rate required for powering SRF cavities. The first method is based on a vector summation of signals of two independently phase-controlled magnetrons. Their output signals are combined by a 3-dB hybrid and the power control is achieved by controlling the phase difference of the rf signals driving each magnetron [13]. The second vector method uses an additional control (modulation) of the phase modulation depth in a single magnetron. Such a modulation results in the rf power being distributed between the fundamental frequency and the sidebands. If the frequency of the depth modulation is much larger than the accelerating field bandwidth, the power concentrated in the sidebands is reflected from the cavity towards a dummy load. Thus, changes of the modulation depth result in a power change at the fundamental frequency [14]. The rf sources intended for powering SRF cavities for both vector methods operate at nominal power and a part of the power is continuously redistributed into a dummy load for absorption. Thus, both vector methods provide an average relative efficiency of about 50%–70% dependent on the range of power control (a few dB) as required for SRF cavities.

The stable operation of a magnetron powered by a voltage below the threshold of self-excitation was observed while studying the possibilities of increasing the efficiency of magnetrons at a wide range of power control. To further explore the observed phenomenon, a simplified analytical model has been developed that considers the phase grouping of the Larmor electrons drifting towards the anode. The grouping is caused by the synchronous wave excited in the magnetron [15]. The model substantiated a novel method of power control in magnetrons over a wide range.

A proof of principle of the novel method providing the highest generation efficiency, low noise, and precise stability was demonstrated in experiments with 2.45 GHz, 1 kW, and CW magnetrons [15]. The method provides a significantly higher average magnetron efficiency at a range of power control up to 10 dB and with wideband phase control. The power control in this technique is realized by variation of the magnetron current over an extended range, where the magnetron, driven by a sufficient (in magnitude) injection-locking signal, may stably operate at a voltage less than the threshold of self-excitation. This provides the extended range of power control as mentioned above. The bandwidth of the power control in this technique is determined by the bandwidth of the current feedback loop in the magnetron high voltage (HV) power supply operating as a current source. Presently the bandwidth may be about of 10 kHz without compromising the efficiency of the power supply. Capabilities of the injection-locked magnetrons for the listed methods of phase and power control have been verified in experiments with 2.45 GHz, CW, 1 kW magnetrons; see Refs. [13–15]. The behavior of magnetrons in the experiments was in accordance with the developed analytic model characterizing the phase grouping and the resonant interaction of the electrons with the synchronous wave in magnetrons. The model allows the estimation of the necessary and sufficient conditions for coherent low-noise generation of magnetrons driven by a resonant injected signal. Consideration of the analytical model in comparison to results of the experiments is discussed.

## II. INTERACTION OF THE DRIFTING CHARGE WITH THE SYNCHRONOUS WAVE IN MAGNETRONS

We consider the simple analytical model, based on the charge drift approximation [16] for conventional magnetrons driven by a resonant rf signal. We discuss a conventional CW magnetron with $N$ cavities ($N$ is an even number) (see Fig. 1), with a constant uniform magnetic field $H$, above the critical magnetic field. The magnetron operates in the $\pi$ mode, i.e., with the rf electric field shifted by $\pi$ between neighboring cavity gaps.

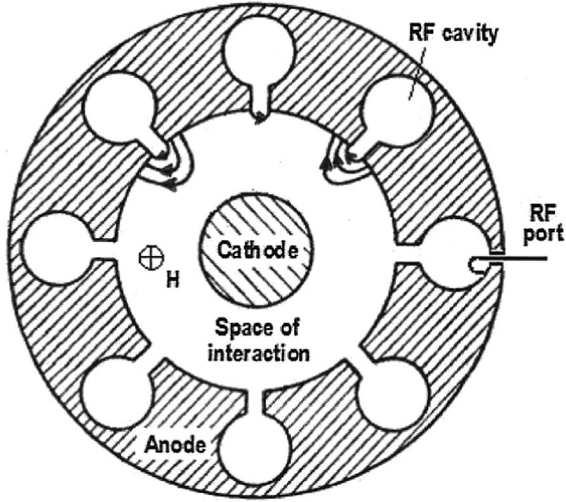

FIG. 1. Schematic sketch of a conventional 8-cavities magnetron. The lines in the space of interaction represent the rf electric field.

We consider the magnetron operating at the frequency $\omega$, being loaded by a matched load with a negligible reflected signal. In the drift approximation, we consider the motion of charge in the center of the Larmor orbit with radius $r_L$, when the Larmor motion of the electron itself is averaged over the cyclotron frequency $\omega$ in the magnetron space of interaction (see [16]). We neglect, as in this reference, the impact of space charge, especially since we consider the CW tubes operating at rather low currents. In this approach, neglecting the azimuthal nonuniformity of the static electric field, the drift of the center of the Larmor orbit with azimuthal angular velocity $\Omega$ in the uniform magnetic field is determined by the superposition of the static electric field, described by the static electric potential $\Phi^0$, and the rf field of the synchronous wave, determined by a scalar potential $\Phi$, which is induced by the magnetron current and the injected resonant rf signal (in a steady state both are in phase). Thus, the drift of the charge can be described in the polar frame by the following system of equations of the first order [17]:

$$\begin{cases} \dot{r} = -\frac{c}{Hr}\frac{\partial}{\partial \varphi}(\Phi^0 + \Phi) \\ \dot{\varphi} = \frac{c}{Hr}\frac{\partial}{\partial r}(\Phi^0 + \Phi) \end{cases}. \quad (1)$$

For the static electric field $\Phi^0 = U \ln(r/r_1)/\ln(r_2/r_1)$ and $E_r = \text{grad}\Phi^0$, $\partial\Phi^0/\partial\varphi = 0$; therefore, $E_\varphi(r) = 0$. Here $U$ is the magnetron feeding voltage; $r_1$ and $r_2$ are the magnetron cathode and anode radii, respectively. The static radial electric field at the magnetron cathode is $E_r(r_1) = U/r_1 \ln(r_2/r_1)$.

We consider a slow rf wave type $\exp[-i(n\varphi + \omega t)]$ excited at the frequency $\omega$ and rotating in the space of interaction with the phase velocity $\Omega = \omega/n$ [17]. The wave number $n = N/2$ sets the same azimuthal periodicity in interaction of the drifting charge with the rf field in the space of interaction. The azimuthal velocity of the wave coincides with the azimuthal drift velocity of the center of the Larmor orbit located on the "synchronous" radius, $r_S$: $r_S = \sqrt{-ncU/[\omega H \ln(r_2/r_1)]}$ [17], ($H < 0$ is assumed).

In magnetrons $r_L \ll 2\pi c/n\Omega$, where the right part of the inequality is the length of the synchronous wave in the magnetron slowing rf system. Therefore, one can consider the interaction of the synchronous wave with an electron rotating along the Larmor orbit as an interaction of the wave with the point charge located in the center of the orbit. This substantiates utilization of the charge drift approximation. The electric field of the synchronous wave has radial and azimuthal components.

Thus, in a conventional magnetron ($\omega \ll \pi c/r_1, \pi c/r_2$), the quasistatic approximation can be used to describe the rotating synchronous wave in the magnetron space of interaction. The scalar potential $\Phi$, satisfying the Laplace equation with inaccuracy $\sim(2\pi r\Omega/c)^2$, [17], for the rotating wave is presented as in Ref. [15]:

$$\Phi = \sum_{k=-\infty}^{\infty} \frac{\tilde{E}_k(r_1) \cdot r_1}{2k} \left[\left(\frac{r}{r_1}\right)^k - \left(\frac{r_1}{r}\right)^k\right] \sin(k\varphi + \omega t), \quad (2)$$

where $\tilde{E}_k(r_1)$ is the amplitude of the $k$th harmonic of the radial rf electric field at $r = r_1$. The form of the potential was chosen so that the azimuthal electric field vanishes at the cathode. The coefficients $\tilde{E}_k(r_1)$ are determined by the requirement to have zero azimuthal electric field at the anode everywhere except the coupling slits of the cavities. The term in the sum of Eq. (2) with $k = n$ has a resonant interaction with the azimuthal motion of the Larmor orbit. As such, we consider only this term. It follows from Eq. (1) that without the synchronous wave, $\dot{r} = 0$ and the drift of charges towards the anode, i.e., magnetron operation, is impossible.

In the coordinate frame rotating with the synchronous wave for $\varphi_S = \varphi + t \cdot \omega/n$ and for an effective potential $\Phi_S$,

$$\Phi_S = U \frac{\ln(r/r_1)}{\ln(r_2/r_1)} + \frac{\omega H}{2nc}r^2 + \frac{\tilde{E}_n(r_1) \cdot r_1}{2n}$$
$$\times \left[\left(\frac{r}{r_1}\right)^n - \left(\frac{r_1}{r}\right)^n\right] \sin(n\varphi_S)$$

one obtains the system of drift equations [16],

$$\begin{cases} \dot{r} = -\frac{c}{Hr}\frac{\partial}{\partial\varphi_S}\Phi_S \\ \dot{\varphi}_S = \frac{c}{Hr}\frac{\partial}{\partial r}\Phi_S \end{cases}, \quad (3)$$

Substituting the potential $\Phi_S$ into Eq. (3) and denoting $\phi_0(r) = \ln\frac{r}{r_1} - \frac{1}{2}(\frac{r}{r_S})^2$ and $\phi_1(r) = \frac{1}{2n}[(\frac{r}{r_1})^n - (\frac{r_1}{r})^n]$, one can obtain the system of drift equations [17,15] in the frame of the synchronous wave expressed via the relative magnitude of the resonant harmonic of the synchronous wave, $\varepsilon$, taken at the cathode:

$$\begin{cases} \dot{r} = \omega \frac{r_S^2}{r} \varepsilon \phi_1(r) \cos(n\varphi_S) \\ n\dot{\varphi}_S = -\omega \frac{r_S^2}{r} \left( \frac{d\phi_0}{dr} + \varepsilon \frac{d\phi_1}{dr} \sin(n\varphi_S) \right) \end{cases}. \quad (4)$$

Here $\varepsilon = \tilde{E}_n(r_1)/E_r(r_1) = \tilde{E}_n(r_1) \cdot r_1 \ln(r_2/r_1)/U$.

The simplified Eq. (4) do not describe the coherent oscillation in a magnetron, but they characterize the resonant interaction of the charge in the center of the Larmor orbit with the synchronous wave.

The top equation in Eq. (4) describes the radial velocity of the moving charge. In accordance with this equation, the drift of the charge towards the anode is possible at $-\pi/2 < n\varphi_S < \pi/2$ with a period of $2\pi$, i.e., only in "spokes." The charge can enter the spoke through the boundaries located at $\pm \pi/2$, [16]. The radial drift velocity is proportional to the synchronous wave magnitude, $\varepsilon$. The condition $\varepsilon \geq 1$ does not allow operation of the magnetron.

The second equation describes the azimuthal velocity of the drifting charge in the frame of the synchronous wave. The second term in the parentheses causes phase grouping of the charge by the resonant rf field via the potential $\phi_1$.

The first term of the equation describes a radially dependent azimuthal drift of the charge, resulting from the rotating frame, with azimuthal angular velocity $-\omega/n$. This term at $r > r_S$ and at low $\varepsilon$ causes the movement of the charge from the phase interval allowed for spokes [15].

Equation (4) were integrated for a typical model of a commercial magnetron described in Ref. [15] with $N = 8$, $r_1 = 5$ mm, $r_2/r_1 = 1.5$, $r_S/r_1 = 1.2$. Considering the charge drifting in the center of the Larmor orbit, we obtained the charge trajectories at $r \geq r_1 + r_L$ for various magnitudes $\varepsilon$ of the rf field in the synchronous wave and during the time interval of the drift of $2 \leq \tau \leq 10$ cyclotron periods allowing coherent contribution to the synchronous wave [16]. The azimuthal boundaries of the charge drifting in a spoke at various $\varepsilon$ obtained by the integration are plotted in Fig. 2(a) assuming a uniform phase distribution of the emitted electrons. The phase interval at $(r_1 + r_L)$ normalized by $\pi$ determines the part of the emitted electrons (in the phase interval admitted for a spoke) contributing to the synchronous wave.

From Fig. 2 it follows that the value of charges in a spoke contributing to the synchronous wave at $\varepsilon = 0.1$ and $\varepsilon = 0.2$ are approximately ¼ and ½, respectively, of the charge's value contributing to the synchronous wave at $\varepsilon = 0.3$ as was shown in [15]. Figure 2(b) shows trajectories of the charges in a spoke at $\varepsilon = 0.3$.

The role of the resonant interaction in magnetrons can be explained in the following manner. In the frame of the slow synchronous wave the rf azimuthal electric field in a spoke can be considered as stationary. The electric field strongly coupled with the resonant mode of the magnetron oscillation acts as a stationary field on the charge drifting in the spoke. This causes the resonant energy exchange between the synchronous wave and the charge. If the azimuthal velocity of the drifting charge is greater than the azimuthal velocity of the synchronous wave, the charge induces oscillation of the resonant mode in the magnetron rf system [18] and contributes to the synchronous wave being decelerated. This increases the amplitudes of the synchronous wave and the rf field in the entire magnetron system. Otherwise, the electric field of the wave accelerates the charge increasing its azimuthal drift velocity. This reduces the wave energy and the amplitude, respectively. The energy exchange of a drifting charge with a synchronous wave causes a variation of the wave radial field amplitude and, as a result, a variation of the azimuthal velocity of the drifting charge [19], ensuring its phase grouping. The increase or decrease of the self-consistent electric field of the synchronous wave resulting from the resonant energy exchange with a charge drifting in a spoke can be determined by the difference in the azimuthal velocities of the drifting charge and the synchronous wave $\Delta v_{AZ}(n\varphi_S, r)$ calculated for the drift. Thus the energy decrement or the energy increment in the synchronous wave are determined by the sign of the $\Delta v_{AZ}(n\varphi_S, r)$ quantity, and one can estimate the value of $\varepsilon$ necessary and sufficient for the coherent generation of the magnetron by integration of $v_{AZ}(n\varphi_S, r)$ over the entire phase interval admissible for a spoke. For more accurate estimations one needs to consider the loss of the drifting charges by migration into the adjacent phase interval.

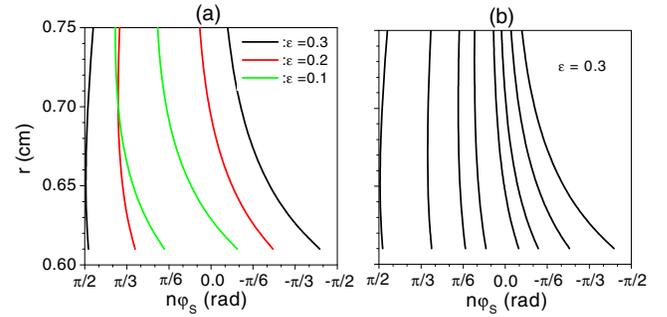

FIG. 2. Phase grouping of the charge drifting towards the magnetron anode in the considered magnetron model. Graph (a) shows the phase boundaries of trajectories of the charges contributing to the synchronous wave in dependence on $\varepsilon$. Graph (b) shows trajectories of the charges in a spoke at $\varepsilon = 0.3$.

Positive values of the integrals of $v_{AZ}(n\varphi_S, r)$ indicate that the rf energy added into the synchronous wave by the drifting charges is larger than the energy removed from the synchronous wave for the phase grouping of the charges. In this case, the magnetron generates coherently. Otherwise, the rf energy generated by the drifting charge and added into the synchronous wave is less than the energy removed from the wave because of the phase grouping. This causes

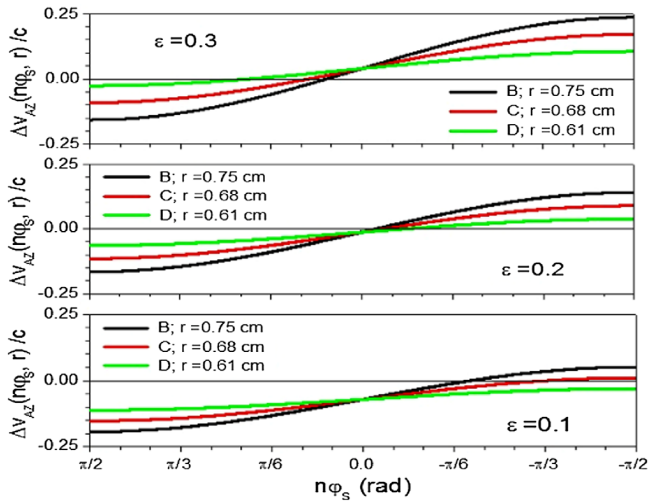

FIG. 3. Azimuthal velocities of the drifting charge and the synchronous wave difference in units of $c$ at various $r$, and $\varepsilon$ vs $n\varphi_S$.

damping of the synchronous wave and disruption of the coherent generation.

Figure 3 represents the phase-dependent function $\Delta v_{AZ}(n\varphi_S, r)$ for various $\varepsilon$ and at various radii of the drifting charge.

As follows from Fig. 3, the quantity $\Delta v_{AZ}(n\varphi_S, r)$ is negative on almost all the radii of the charge trajectory over the entire phase interval admissible for the spoke at $\varepsilon = 0.1$. This indicates a decrement of the energy of the synchronous wave causing its damping. The magnetron cannot operate at so low value of $\varepsilon$. At $\varepsilon = 0.2$ the decrement of the energy of the synchronous wave also dominates. The increment at $-\pi/24 \leq n\varphi_S \leq -\pi/2$ [where the values of $\Delta v_{AZ}(n\varphi_S, r)$ are positive] cannot compensate the wave decrement causing damping of the synchronous wave. The graphs in Fig. 3 indicate that the coherent generation of the modeled magnetron at $\varepsilon \leq 0.2$ is impossible while $\varepsilon \cong 0.3$ allows stable operation of the considered magnetron model. It means that the radial component of the rf field of the synchronous wave on the cathode which is approximately 30% of the static electric field on the cathode provides stable operation of magnetrons. In [15] it was shown that the value $\varepsilon \cong 0.3$ corresponds to a resonant injected signal power of about $-10$ dB of the magnetron nominal power. As is shown below, such a resonant injected signal provides advanced capabilities of magnetrons.

The resonant interaction of a synchronous wave with the drifting charges, which are appropriately grouped in phase by the synchronous wave in a magnetron, provides the energy exchange between the electron flow and the self-consistent rf field resulting in coherent generation of the tube. This is the basic principle of magnetron operation. In this, the principle of the operation of magnetrons is similar to the principle of the operation of such coherent generators as FEL, etc.

As a result of the phase grouping, the graphs in Fig. 2 indicate violation of the initial uniform distribution of the azimuthal velocities of the drifting charges.

We note once again that the energy exchange between the synchronous wave and the averaged motion of Larmor electrons in magnetrons is considered in the drift approximation to be the energy exchange between the wave and charges drifting in the centers of Larmor orbits.

At low $\varepsilon$, the loss of charges coherently contributing to the synchronous wave, Fig. 2(a), is equivalent to a reduction of the magnetron current in the spokes, i.e., the magnitude of the induced synchronous wave. Also, at low $\varepsilon$, part of the charge migrates from the spoke into the adjacent phase interval [15], where it is accelerated by the synchronous wave. This also reduces the synchronous wave magnitude, deteriorating phase grouping. Thus, for stable operation of the magnetron, one needs to prevent a decrease of $\varepsilon$, otherwise the insufficient magnitude of the synchronous wave will cause noisy operation of the tube and, in principle, may lead to disruption of generation resulting from loss of coherency as is shown below.

Moreover, the acceleration of the lost charge increases the Larmor radius of the rotating electron which may hit the cathode increasing the magnetron cathode losses. For the typical magnetron model [15], the losses of the drifting charge coherently contributing to the synchronous wave are minimized at $\varepsilon \sim 0.3$ in the main part of the charge drift trajectories. Thus, a sufficient magnitude of the synchronous wave protects from loss of coherency and prevents an increase of the cathode losses even at low magnetron power.

A resonant injected signal driving the magnetron in accordance with the energy conservation law increases the rf energy stored in the magnetron cavities and in the interaction space. Since the rf energy in the magnetron is determined by the static electric field, the injected resonant signal is equivalent to an increase of the magnetron feeding voltage. Thus, a sufficient power level of the injected resonant signal allows the magnetron to start up even if the magnetron feeding voltage is somewhat less than the threshold of self-excitation. In this case the magnetron current should be less than the minimum current when the tube is self-excited [15] (the free run operation). This allows stable operation of the magnetron over an extended range of current (power) control. A lack of rf voltage in the synchronous wave induced by the lower magnetron current is compensated by the injected resonant signal providing stable operation of the tube. As follows from the charge drift model estimations and experiments with the 2.45 GHz, 1 kW magnetrons, the injected resonant signal with a power level of about $-10$ dB of the nominal magnetron power allows magnetron power control over the range of 10 dB by deep variation of the magnetron current [15]. Operation of the magnetron at a current less than the allowable minimum current for free run operation in accordance with its volt-amp (V-I) characteristic is possible at a magnetron voltage below the threshold of self-excitation. At a sufficient power of the resonant driving signal, the loss of the drifting charge and the cathode losses remain low even at small magnetron currents. Thus, one can expect the highest efficiency from a magnetron driven by a sufficient resonant signal providing stable operation of the tube in an extended range of current (power) control.

## III. IMPACT OF MAGNITUDE OF THE INJECTION-LOCKING SIGNAL ON THE MAGNETRON OPERATION

How the magnitude of the injected resonant signal affects the magnetron operation was studied mainly with a 2.45 GHz, 1.2 kW magnetron type 2M137-IL with a permanent magnet operating in the CW regime [15]. The measurements were performed at the nominal voltage of the magnetron filament.

The magnetron was started up and frequency locked by the HP 8341A generator via a solid-state amplifier (SSA) and 36.6 dB traveling wave tube (TWT) amplifier providing CW locking power up to 100 W. The magnetron was powered by an Alter switching high voltage (HV) power supply type SM445G with a current feedback loop, operating as a current source and allowing current control. The setup was used to measure the magnetron power and efficiency, the noise at various levels of power of the magnetron and locking signal, and the stability of the carrier frequency of the tube.

The V-I characteristic of the magnetron, Fig. 4, was measured in the CW regime. The magnetron cathode high voltage and current have been measured by the calibrated compensating divider and the transducer, respectively, via an oscilloscope. The inaccuracy of calibrations did not exceed ±1%.

Start up and injection locking of the magnetron at the current less than the minimum current in free run operation was provided by injection of the resonant signal with power $P_{Lock} = 100$ W. Figure 4 shows the V-I characteristic with standard deviation error bars.

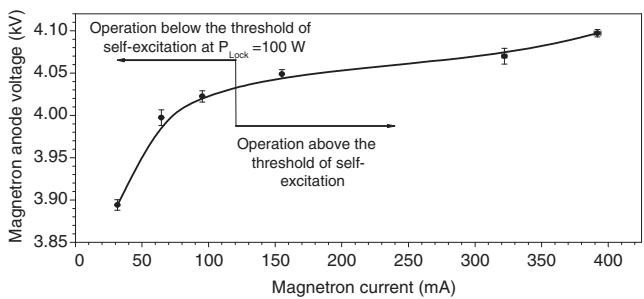

FIG. 4. Magnetron V-I characteristic measured at $P_{Lock} = 100$ W. The solid line (B-spline fit) shows the available range of current with stable operation of the tube at power of the resonant injected signal $P_{Lock} = 100$ W.

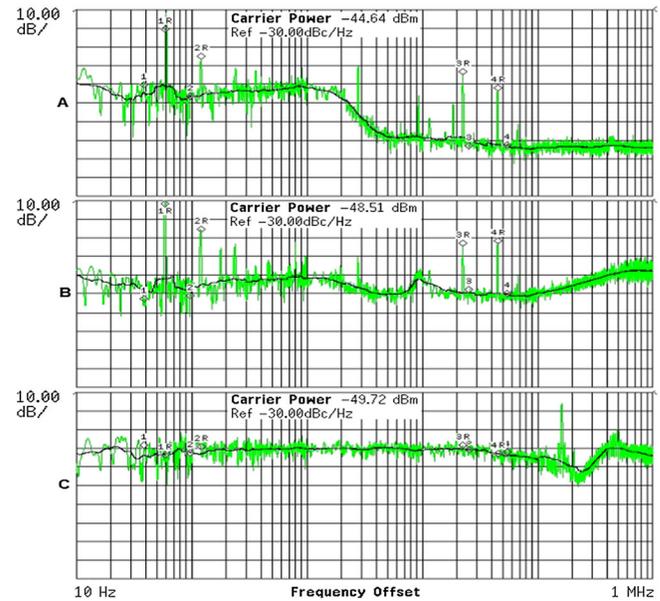

FIG. 5. Spectral power density of the noise at various power levels of the locking signal at $P_{Mag} = 100$ W. Traces A, B, C show the spectral power noise density at $P_{Lock} = 100$ W, 30 W, and 10 W, respectively. Black traces show the averaged spectral power density of the noise.

As was discussed above, a power level of the injected resonant signal of about −10 dB of the magnetron nominal power is sufficient to start up the tube, providing stable operation at currents over the range of 31.6-392 mA, i.e., at the range of power control ≥10 dB.

A decrease of the resonant injected signal when the magnetron power is less than minimum power admitted for free run operation decreases the $\Delta v_{AZ}(n\varphi_S, r)$ values. It increases noise and may lead to loss of coherency, Fig. 3, and may stop generation. A decrease of the resonant injected signal at the nominal magnetron power increases relative fluctuations of the synchronous wave magnitude.

We studied these phenomena in two sets of experiments [15], measuring the spectral density of the power of the magnetron noise relative to the power of the carrier frequency; see Figs. 5 and 6.

In the first set, we studied the operation of the magnetron below the threshold of self-excitation at an output power of 100 W, Fig. 5, when $\varepsilon$ is determined mainly by the injected resonant signal and one can assume that the fluctuations of $\varepsilon$ are quite small due to the stability of the injected resonant signal. Fed below the Hartree voltage, the magnetron being started up and injection locked at various powers of the locking signal shows a dramatic increase of the measured density of noise power in the range from 1 kHz to 1 MHz if power of the injected resonant signal is decreased from 100 W to 10 W; see Fig. 5, traces A, B, C.

The dramatic increase of the noise power density beginning at 1 kHz or less (this is much less than the switching frequency of the HV power supply) can be explained as relaxation oscillations with a characteristic frequency of $f_c \sim 1/(Z_S \cdot C_c)$ resulting from partial or total loss of coherency in magnetron oscillations at insufficient $\varepsilon$ values, traces B and C, respectively. Here $Z_S$ is the magnetron static impedance and $C_c$ is the capacitance of the magnetron cathode HV circuitry. The relaxation oscillations are caused most likely by a short overvoltage in the magnetron resulted from a disruption of the coherent oscillation. This may restore conditions for a start up of the tube with an appropriate phase grouping during a short time. For the magnetron operating with a power of 100 W, $Z_S$ is about 100 kOhms; see Fig. 4. This corresponds to $f_c \geq 1$ kHz at $C_c \leq 10$ nF. Note that the repeating sidebands in the magnetron noise spectra in Figs. 5 and 6 result from the magnetron switching power supply and switching power supply of the TWT amplifier. The sidebands caused by the TWT switching power supply are seen in Fig. 6, trace D, showing the spectral density of the noise power of the injection-locking signal when the magnetron high voltage was off.

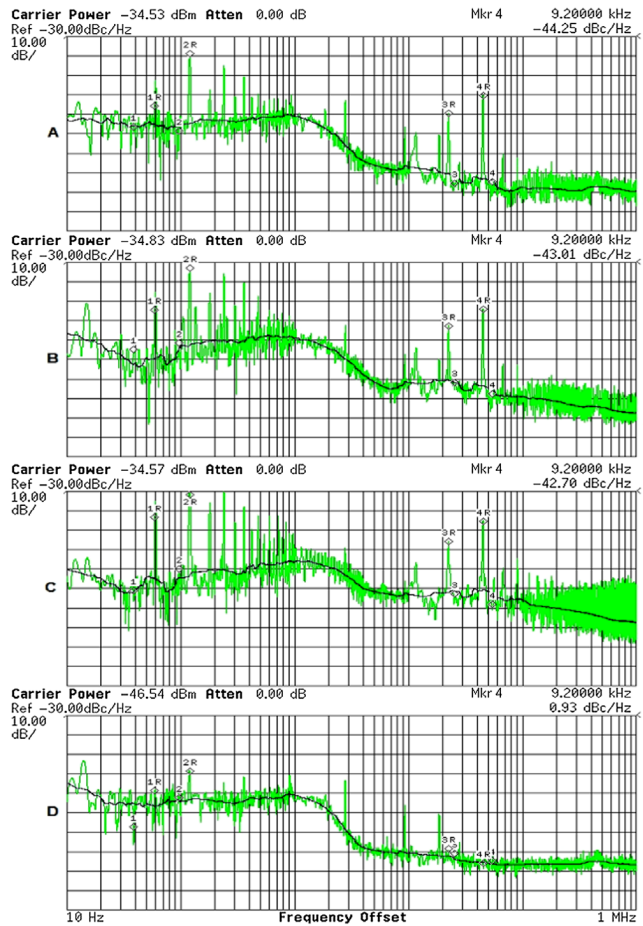

FIG. 6. The spectral power density of the noise of magnetron at the output power of 1 kW, at the power of the locking signal of 100, 30, and 10 W, traces A, B, and C, respectively [15]. Traces D are the spectral power density of noise of the injection-locking signal ($P_{\text{Lock}} = 100$ W), when the magnetron feeding voltage is off. Black traces show the averaged spectral power density of the noise.

Measured spectral power densities of the noise vs. power of the locking signal at the nominal magnetron power, when the injected resonant signal affects the resonant interaction less, are plotted in Fig. 6.

Since the fluctuations of $\varepsilon$ are approximately proportional to the fluctuations of the charge reaching the anode, at low locking power they cause notable amplitude modulations of the synchronous wave and the modulation sidebands with frequencies $f_M$, which are integrated by the magnetron cavity with the loaded Q-factor, $Q_M$, over a time $\tau_M \sim Q_M/(\pi \cdot f_M)$. Thus the noise caused by amplitude fluctuations in the synchronous wave is limited to the MHz range for 2.45 GHz magnetrons. Note that the nonresonant sidebands do not interact resonantly with the drifting charge.

The traces C-A in Fig. 6 illustrate a notable reduction of the magnetron spectral power density of the noise (by ~20 dBc/Hz) at frequencies >100 kHz when the power of the injection-locking signal is increased from 10 W to 100 W. This indicates a reduction of fluctuations of the rf field in the entire magnetron system caused by an increase of the stable injection-locking signal.

Comparison of traces B and C in Fig. 5 with traces B and C in Fig. 6, measured at the same power of the locking signal ($P_{\text{Lock}} = 30$ W and 10 W, respectively) demonstrates a significant difference in the spectral power density in the low-frequency range (<100 kHz) between low output power (100 W) and high output power (1 kW) of the magnetron.

In the frequency range of (1–100 kHz), Fig. 6, traces A-C, the decrease of the locking power from −10 dB to −20 dB increases the power density of noise by ~20 dBc/Hz. This indicates deterioration (partial losses) of coherency at the almost nominal power of the tube driven by the insufficient injection-locking signal. Disruption of coherent generation of the magnetron caused by total loss of coherency was observed in operation below the threshold of self-excitation and low locking signal [Fig. 5(c)]. It was not observed at the almost nominal magnetron power and the low power of locking signal; see Fig. 6(c).

Thus, the noticeable fluctuations of the charge contributing to the coherent generation cause fluctuations in the magnitude of the synchronous wave resulting in noise in the MHz range. The noise is effectively suppressed by increasing the injection-locking signal to −10 dB.

At a magnetron voltage somewhat less than the threshold of self-excitation, the tube power is low and the magnetron demonstrates loss of coherency at an insufficient injected resonant signal. This is manifested in a dramatic increase of noise in the low-frequency range resulting from a significant decrement of the synchronous wave energy, traces B, and a total disruption of coherent oscillation, traces C, respectively in Fig. 5.

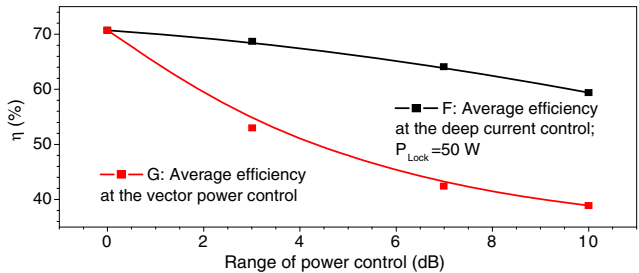

FIG. 7. Averaged magnetron efficiency vs range of power control for various methods of control. Curve F shows the average efficiency of the 1.2 kW magnetron driven by the injected resonant signal of −10 dB and measured at deep magnetron current control. Curve G shows the average efficiency of 1 kW magnetrons with vector power control [12,13].

At the injected resonant signal of −10 dB the magnetron demonstrates a noise power density of −100 dBc/Hz or less in the frequency range of 1 MHz avoiding loss of coherency for an output power ranging from 100 W to 1000 W; see Figs. 5 and 6, traces A.

The magnetron absolute efficiency, $\eta$, was determined as the ratio of the measured magnetron rf power to the output power of the magnetron HV power supply at various values of the magnetron rf power [15]. The dependence of the magnetron average efficiency on the range of power control (in dB) assuming a linear variation of the magnetron power in the given range is shown in Fig. 7.

The graphs demonstrate the highest average efficiency of the magnetron driven by a sufficient resonant injected signal at a wide range of (current) power control in comparison with the vector power control methods.

Precise stability of the carrier frequency during operation of the magnetron in the regime with wide-range power control at a sufficient locking signal is demonstrated in Fig. 8 [15], which shows the offset of the carrier frequency at various power levels of the magnetron and the locking signal.

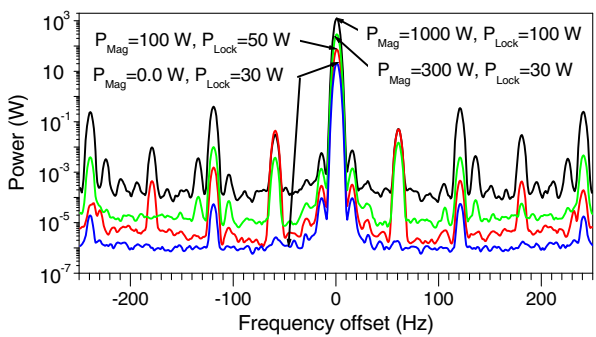

FIG. 8. Offset of the carrier frequency at various power levels of the magnetron, $P_{\text{Mag}}$, and the locking signal, $P_{\text{Lock}}$. The trace $P_{\text{Mag}} = 0.0$ W, $P_{\text{Lock}} = 30$ W shows the frequency offset of the injection-locking signal.

The measured offset of the carrier frequency does not demonstrate any broadening of the magnetron spectral line over a wide range of the magnetron power control. This implies that no noticeable noise is produced in the magnetron when operating below the threshold of self-excitation at a sufficient driving resonant signal and thus no losses of coherency are encountered.

Trace $P_{\text{Mag}} = 0.0$ W, $P_{\text{Lock}} = 30$ W in Fig. 8 was measured with the magnetron high voltage turned off.

The low frequency sidebands in all traces are caused by 60 Hz modulation in the switching power supplies of the magnetron and TWT amplifier.

## IV. A WIDEBAND PHASE CONTROL IN INJECTION-LOCKED MAGNETRONS

The wideband phase control required for an SRF cavity powered by a magnetron rf source was first studied in detail using the phase modulation technique with various configurations of transmitters in the pulsed regime with CW, 2.45 GHz magnetrons [13]. A pulsed modulator could power the two magnetrons; it used a partial discharge of a storage capacitor providing pulse duration of 5 ms, with voltage droop of about 0.4% and a negligible ripple. Experiments were performed with single and 2-cascade magnetrons injection locked by the phase-modulated signal. Two magnetrons with a frequency offset of ≈4.7 MHz and a power of about 0.5 kW have been used in experiments. Each one was installed in a separate module.

The single injection-locked magnetrons were tested in a configuration using the module with the magnetron locked by the CW TWT amplifier and fed by the modulator, while the other module was disconnected from the amplifier and the modulator. The 2-cascade magnetron was tested in a configuration in which the magnetron in the first module was injection locked by the TWT amplifier while the magnetron in the second module was connected via an attenuator to the first module rf output. In this case, the magnetron in the second module was injection locked by the pulsed signal of the first magnetron, lowered in the attenuator. Both magnetrons were fed by the same modulator. Experiments demonstrated operation of the 2-cascade magnetron, injection locked at the average of the offset frequencies, at attenuator values in the range of 9–20 dB [13].

A study of the wideband phase control of the single and 2-cascade injection-locked magnetrons was performed using internal phase modulation by a harmonic signal in the N5181A generator. The broadband (2–4 GHz) TWT amplifier did not distort the phase-modulated signal, injection locking the magnetrons.

The transfer function magnitude characteristics (magnitude Bode plots expressing the magnitude response of the device as a function of the frequency of the phase modulation) averaged over 8 pulses of the single and 2-cascade magnetrons, Fig. 9, were measured in the phase modulation domain by the Agilent MXAN9020A signal analyzer at various power levels of the injection-locking signal.

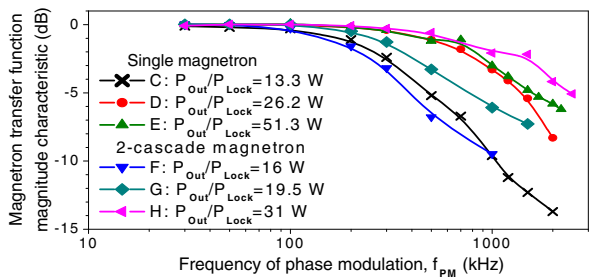

FIG. 9. Transfer function magnitude characteristics of the phase control measured in the phase modulation domain with single and 2-cascade injection-locked magnetrons at various power levels of the injection-locking signal, $P_{\text{Lock}}$.

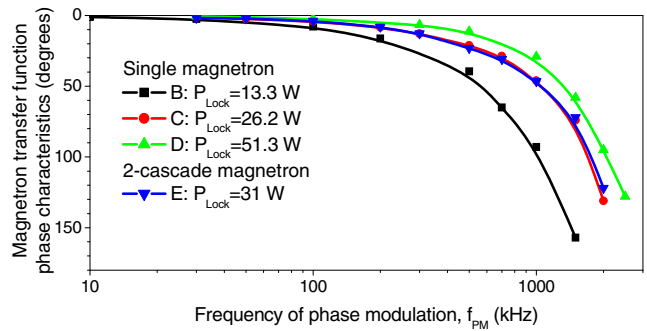

FIG. 10. Transfer function phase characteristics of single and 2-cascade magnetrons injection locked by the phase-modulated signal vs the modulating frequency, $f_{\text{PM}}$.

The measurements have been performed at a magnitude of the phase modulation of 0.07 rad., and at a magnetron output power of $P_{\text{Out}} \approx 450$ W [13]. The measurements with the 2-cascade magnetron were carried out at an attenuator value of ≈13 dB, i.e., the second cascade (tube) operated at $P_{\text{Lock}} \approx 25$ W. In Figs. 9 and 10 in the plots concerning the 2-cascade magnetron, $P_{\text{Lock}}$ denotes the power of the signal locking the first cascade.

The phase response of the magnetrons to the phase-modulated signal vs frequency of the phase modulation, $f_{\text{PM}}$, has been measured with the calibrated phase detector including the phase shifter, double balanced mixer and low pass filter, at a magnitude of modulation of 0.35 rad, and a magnetron output power of $P_{\text{Out}} \approx 500$ W [13]. The transfer function phase characteristics (phase Bode plots) for the single and 2-cascade injection-locked magnetrons at various power values of the locking signal are shown in Fig. 10. The plots consider the magnitude response of magnetrons to the phase-modulated locking signal and the phase detector instrumental function.

In accordance with the standard criteria (−3 dB for the magnitude characteristics and 45° for the phase characteristics) Figs. 9 and 10 demonstrate an allowable bandwidth of the closed feedback loops exceeding 1 MHz at $P_{\text{Lock}} \approx$ 30 W for a single 2.45 GHz, 1 kW magnetron. The estimates of the bandwidth obtained from the magnitude and phase characteristics are in agreement.

The plots in Fig. 10 show that the phase deviation of the output magnetron signal from the phase of the injection-locking signal becomes noticeable beyond $f_{\text{PM}} \sim 10$ kHz and increases with larger frequencies of the phase modulation or lower power of the locking signal. It indicates that the phase-locking concept is applicable in a relatively narrow bandwidth of the phase control.

Expressing the ratio of the magnetron (nominal) power to the power of the resonant driving (injection-locking) signal as $G = P_{\text{Nom}}/P_{\text{Lock}}$ one can say that the 1 MHz bandwidth of the closed feedback loops for the 2.45 GHz magnetrons is provided by $G \sim$ 10–13 dB per cascade (tube).

The vector methods of power control in magnetrons are reduced to phase control by the phase-modulated injection-locking signal. The vector method of managing the depth of the phase modulation to control the magnetron power was tested by a single-cell 2.45 GHz SRF cavity with a loaded Q-factor $Q_L \approx 2 \times 10^7$, powered by the single CW magnetron type 2M137-IL. The measured rms phase and amplitude deviations of the accelerating field in the cavity at 4K did not exceed 0.26° and 0.3%, respectively, at a closed feedback-loop bandwidth of ∼100 kHz [14]. For 650 MHz SRF cavities with R/Q = 610 Ohms the loaded Q-factor, $Q_L \approx 2 \times 10^7$ corresponds to an accelerated current of ∼0.5 mA. A higher beam current decreases $Q_L$ because of broadening of the accelerating mode bandwidth. Then the bandwidth of the closed feedback loop can be much narrower; e.g., at a beam current of 10 mA ($Q_L \sim 3 \times 10^6$), the required feedback loop bandwidth of ≈4 kHz is sufficient to suppress amplitude modulations caused by microphonics with the frequency cutoff of 65 Hz to the level of 0.3%.

## V. MODELING OF A DYNAMIC POWER CONTROL BY MANAGEMENT OF MAGNETRON CURRENT

The capability of the proposed method for deep dynamic power control was verified by a modulation of the magnetron current via control of the HV switching power supply within a current feedback loop; see Fig. 11 [20].

The low-frequency harmonic analog signal driving the control input of the SM445G power supply regulated the current of the magnetron driven by the injected resonant rf signal at $P_{\text{Lock}} = 100$ W. The magnetron power was determined by measurements of the magnetron current with a calibrated transducer and by measurements of rf power with a calibrated detector vs the magnetron current. The inaccuracy of the calibration and the rf power measurements did not exceed ±1%. The traces were averaged over 16 runs reducing the noise caused by operation of the switching power supply. The bandwidth of the power control was limited by the bandwidth of the power supply feedback loop since the power supply was designed for slow current control. Nevertheless, the harmonic shapes of the measured traces demonstrate a proof of the principle of wide range dynamic power control by a wide-range management of the magnetron current.

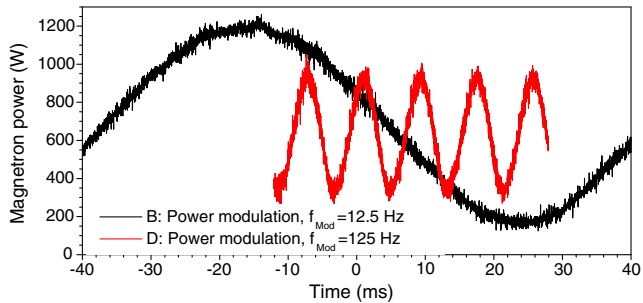

FIG. 11. Modulation of the magnetron power managing the magnetron current by a harmonic signal controlling the SM445G HV switching power supply.

## VI. CONCEPTUAL SCHEME OF A HIGHLY-EFFICIENT HIGH-POWER MAGNETRON TRANSMITTER

A conceptual scheme of a single-channel transmitter based on magnetrons, allowing dynamic wideband phase and midfrequency wide-range power management at the highest average efficiency is shown in Fig. 12 [20].

In this scheme the first, low-power magnetron, provides phase modulation (control) of the signal frequency locking the second, high-power magnetron. Power control in the required range (up to 10 dB) is realized by modulation (control) of current in the high-power magnetron which at low current operates at a voltage less than the threshold of self-excitation. Stable and low-noise operation of the high-power magnetron driven by a resonant injected signal at a voltage less than the threshold of self-excitation is provided by the injected signal with sufficient power.

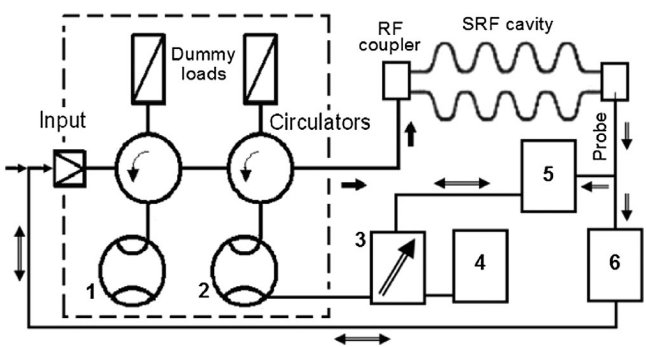

FIG. 12. Conceptual scheme of a 2-cascade magnetron single-channel transmitter allowing dynamic phase and power control at highest efficiency and the rates required for high-current accelerators. 1: low power magnetron; 2: high-power magnetron; 3: low-power HV power supply controlled within the feedback loop; 4: main uncontrolled HV power supply; 5: current/voltage controller within the low level rf (LLRF) system for the controlled HV power supply; and 6: phase controller within the LLRF system.

The current management may be provided by a high-voltage system including a low-power switching power supply within a current feedback loop providing a controllable voltage in the range of a few percent of the nominal magnetron voltage. The controllable power supply is biased by an uncontrollable HV power supply so that the magnetron total feeding voltage can be varied from the nominal value to the value below the threshold of self-excitation which is a few percent less than the magnetron nominal voltage. Presently the bandwidth of such a magnetron power (current) control can be ∼10 kHz without compromising the combined power supply efficiency as was noted above. This bandwidth is sufficient for various SRF accelerators with beam currents >10 mA, e.g., for ADS-class projects. Thus, a single-channel magnetron transmitter with a sufficient injection-locking signal (about −10 dB), as substantiated by the presented experimental results and the modeling, will allow the required dynamic wideband phase control and power control in the midfrequency range in ADS-class projects or other high-current accelerators.

The control (modulation) of the magnetron current causes phase pushing in the frequency-locked magnetron. At a bandwidth of the phase control in the MHz range, one expects the phase pushing to be eliminated to a level less than −50 dB, suitable for various SRF accelerators.

The 2-cascade transmitters based on the injection-locked magnetrons provide bandwidth of phase control up to the MHz range at a G value of about 20–25 dB, i.e., 10–12 dB per cascade; see Figs. 9 and 10. As was demonstrated above and in Ref. [15], this value is also sufficient for operation of a single-channel 2-cascade magnetron transmitter with power control in the range up to 10 dB by control of the magnetron current over an extended range.

Thus a transmitter with the conceptual scheme presented above, satisfying requirements for various superconducting high-current accelerators, will provide the highest efficiency at the lowest capital and operation costs. The highest efficiency is very important for ADS class and other projects of GeV-scale superconducting accelerators of megawatt beams.

The developed technique of magnetron power control by the deep variation of the magnetron current can be combined with the vector methods of power control. In this case the transmitter will provide wideband phase and power control at the highest efficiency.

## VII. SUMMARY

A simplified analytical model recently developed for explanation and substantiation of the novel technique for power control in efficient, high-power magnetron rf sources intended for GeV-scale MW-class accelerators was used for consideration and analysis of the resonant interaction of the Larmor electrons with a slow synchronous wave in magnetrons driven by a resonant injected signal. The analysis considers the coherent generation of magnetrons powered below and above the threshold of self-excitation and driven by the resonant injected signal to be a result of the resonant interaction of a synchronous wave with Larmor electrons, which are appropriately grouped in phase by the wave.

The analysis demonstrates that a sufficient strength of the radial rf field of the synchronous wave on the cathode ($\varepsilon \approx 0.3$) causes an appropriate phase grouping of the drifting charge and the energy exchange between the drifting charges and the synchronous wave resulting in the coherent contribution to the wave. This provides the necessary and sufficient conditions for coherent generation of the magnetron. The analysis allows an estimate of the necessary and sufficient conditions for the coherent generation and explains the impact of the resonant injected signal on stability, efficiency, and noise of magnetrons, all in agreement with experiments. Thus the analytical model representing the resonant interaction of the electron beam with the slow synchronous wave in magnetrons reflects the basic principle of magnetron generation and reflects the phase grouping as the basic property of magnetron operation.

The resonant interaction analysis (model) of the coherent generation of the magnetron driven by the resonant injected rf signal and fed below the threshold of self-excitation at the extended range of power control is well verified in the experiments. Presentation of the measured phase and magnitude characteristics of the magnetron transfer function allows one to choose the bandwidth of the feedback loops for the control of magnetrons with rates required for superconducting modern high-current accelerators. Based on the obtained results, the concept of the rf source for MW-class superconducting accelerators is proposed.

## ACKNOWLEDGMENTS

The work was supported by Fermi Research Alliance, LLC under Contract No. De-AC02-07CH11359 with the US DOE, Office of Science, Office of High Energy Physics, and collaboration Muons, Inc.—Fermilab. We are very thankful to Mr. B. Chase and Mr. R. Pasquinelli for help in experiments, Dr. Ya. Derbenev for stimulation of this work, and Dr. R. Thurman-Keup, Dr. F. Marhauser, Dr. V. Balbekov, and Dr. Yu. Eidelman for fruitful discussions. We are indebted to Dr. Helen Edwards for her support of this work and we honor her memory. Without her deep interest and insight into the physics of the interaction of particles and waves, this work would hardly have been completed.